\documentclass[11pt]{article}

\begin{document}\thispagestyle{empty}
%\begin{flushright}
%\framebox{\small BRX-TH~542}\\
%\end{flushright}

\vspace{.8cm} \setcounter{footnote}{0}

\begin{center}{\Large{A Comment on hep-th/0608078}}\\[8mm]

S. Deser\\
Department of Physics, Brandeis University\\ Waltham, MA 02454,
USA \\ and \\
  Lauritsen Laboratory, California Institute of Technology \\ Pasadena, CA 91125, USA.
\vspace{.4cm}

{\small(\today)}\\[1cm]
\end{center}

\begin{abstract}
\noindent The above hep-th posting purports -- erroneously -- to be a comment on a Note
by me in gr-qc.  
\end{abstract}

\subsubsection*{}
In a brief pedagogical Note [1], I discussed some  pitfalls in expressing
D=3 models in lower-order (Palatini) form.
In the cited submission, the authors instead construct a baroque, but {\it not}
lower-order, reformulation of a subcase,
free-field topologically massive electrodynamics [2].
The above perhaps permits a comment on the authors' previous six papers -- all
on $D=2$ gravity: had they read my [3],
their efforts might have been averted.

\subsubsection*{}
I thank R. Jackiw for (unlike these authors) advising me of their hep-th submission.  My work was supported by NSF Grant PHY 04-01667.

\vskip 1cm
\noindent [1] S. Deser, gr-qc/0606006; Class. Quantum Grav. 23 (5773)2006.

\noindent [2] S. Deser, R. Jackiw \& S. Templeton, Ann. Phys. 140(372)1982.

\noindent [3] S. Deser, gr-qc/9512022; Found. of Phys. 26(617)1996.

%\section*{References}

%\bibliography{DFT}

 \end{document}